# Green Internet of Things: The Next Generation Energy Efficient Internet of Things

Navod Neranjan Thilakarathne[1][0000-0002-7761-2035], Mohan Krishna Kagita[2] and W.D Madhuka Priyashan[2]

[1] Department of ICT, Faculty of Technology, University of Colombo, Srilanka
[2] School of Computing and Mathematics, Charles Sturt University, Melbourne, Australia
[2] Department of Mechanical and Manufacturing Engineering, Faculty of Engineering, University of Ruhuna, Galle, Sri Lanka
`navod.neranjan@ict.cmb.ac.lk`, `mohankrishna4k@gmail.com`,
`madhuka.p@mme.ruh.ac.lk`

**Abstract.** The Internet of Things (IoT) is seen as a novel technical paradigm aimed at enabling connectivity between billions of interconnected devices all around the world. This IoT is being served in various domains, such as smart healthcare, traffic surveillance, smart homes, smart cities, and various industries. The main functionality of IoT includes sensing out the surrounding environment and collect data from the surrounding and transmit those data to the remote data centers or to the cloud. This sharing of vast volumes of data between billions of IoT devices generates a large demand for energy and increases energy wastage in the form of heat. The idea of reducing the energy consumption of IoT devices and keeping the environment safe and clean is envisaged by the Green IoT. Inspired by achieving a sustainable next generation IoT ecosystem and guiding us towards making a healthy green planet, we first offer an overview of Green IoT (GIoT), and then the challenges and the future directions with regards to the GIoT are presented in our study.

## 1 Introduction

Day by day IoT related technologies are getting close to our lives in various forms. It is believed that IoT will become a revolutionizing technology that can change the phase of our world [1,2,12,23-25,27]. This IoT is capable of facilitating the connection of billions of digital devices. It can also be known as the advanced version of Machine to Machine (M2M) communication, where each machine or a digital object communicates without human interference with another machine or a digital object [1,3,10]. IoT and related technologies encompass a variety of devices such as various actuators, sensors, network gateways, and mobile devices that are linked through the Internet and these things or objects can sense the environment, transfer information, and interact with each other in variety of ways [1,3,4]. The IoT devices like any other device utilize energy to function and operate, but in some cases, these devices utilize more than the required energy which leads to waste of energy by generating unnecessary heat. This waste of energy and unnecessary heat should be reduced for the benefits of economy and the safety of our environment. Due to the latest technological advancement and the exponential growth of working IoT devices, it also increased the energy demand that is required for device functionality. This created the desire of low power consuming IoT also known as GIoT [2,12]. It is claimed that by adapting to low energy consumption strategies, IoT will play a key role in mitigating the climate crisis in forthcoming years [40].

Owing to the wide scale of digital context and the energy consumption of wide variety of energy-hungry digital devices, energy consumption rates have reached distressing levels [1,3,4,8,36]. With the increased use of connected IoT devices and the amount of data that is generated and transmitted across the IoT infrastructure, scientists expect a tremendous data rate and a huge content size at the price of an exceptional carbon emissions into the environment [1,40]. The latest report [18], has shown that emissions of carbon dioxide from cellular networks will be 345 million tons by 2020 and are projected to rise on an annual basis. As a result, due to these massive carbon dioxide emissions and the environmental and health challenges, clean or green technologies are becoming an enticing research field [1]. On the other hand, this GIoT ecosystem itself faces various challenges such as security and quality of service concerns, the complexity of adopting a universal architecture, heterogeneous devices, and so on. As a result, researchers and industry are working towards developing novel solutions such as innovative GIoT solutions and integrating with enabling technologies like cloud computing, fog computing, and so on to overcome such challenges. The main objective of this work is to provide readers with a brief understanding of GIoT and utilizations of GIoT techniques towards achieving an eco-friendly sustainable world. For the remaining part of this paper, section two provides an overview of the architecture of IoT and various applications of IoT. In section three we discuss GIoT and the approaches for achieving GIoT. The challenges and the future directions are discussed in section four and finally, the paper is concluded in section five.

## 2    Architecture and the Applications of IoT

### 2.1    Architecture of IoT

Many researchers have proposed different IoT architectures. But there is no single unit of IoT architecture that is generally agreed upon [2,7,14]. The most well-known IoT architecture comprises three layers namely the perception, network, and the application layer [19-22,39], as depicted in Fig 1.

- The perception layer comprised of physical IoT devices that consists of various actuators and sensors for sensing the environment and collecting information.
- The network layer supports the transmission and the processing of sensor data gathered by the perception layer. It is mainly used for connecting to other smart things, network devices, and servers.

- The application layer holds the responsibility for supplying the user with application-specific services. It consists of various applications that facilitate smart cities, smart homes, smart healthcare, and other IoT domains.

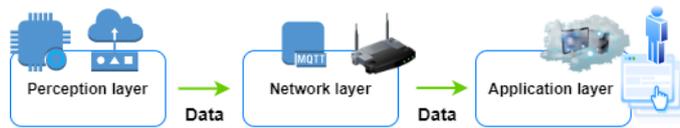

**Fig. 1.** Architecture of IoT.

The functionality of IoT comprised of several key stages that is the identification, sensing, communication, computing, services, and semantics [14,26]. The identification stage ensures that the information or service required reaches the correct address. Sensing deals with the collection of information from different resources and sending this information to the datacenters or to the cloud. Various sensing attributes such as air quality, humidity, temperature and so on can be collected by these IoT sensing devices. IoT communication enables IoT devices to provide specific services for users and most of the time it is carried out using wireless media such as Bluetooth, BLE (Low Energy Bluetooth), and Wi-Fi, etc. Various microcontrollers, microprocessors, and many software applications perform computations. Depending on the context and domain in which the IoT devices reside, services can vary and provide a variety of services for end-users. Finally, Semantics deals with the gathering of intelligent knowledge to make quality decisions.

## 2.2 Applications of IoT

It is no doubt that, by monitoring different situations and making smart decisions to optimize our lifestyle, IoT is being revolutionizing our daily lives. Following we discuss few application domains of IoT [1,3,4,7,14,27,28,35,39-41].

- Smart Homes: Integrating the home environment with various IoT devices and technologies like smart TVs, home security systems, heating systems can facilitate tracking the activities of inhabitants and controlling the home environment [11,12].
- Food Supply Chains (FSC): Integrating IoT technologies in the supply chain allows the vendors to keep track of their products, from the farm to the consumers [12,17].
- IoT in Mining Industry: IoT technologies can be used to ensure the safety of miners and can provide valuable information regarding the mining process to the mining companies. In addition, it facilitates communication and allows companies to track down the location of miners.
- IoT in Transportation: IoT allows to track vehicles and products using various tracking devices. (e.g.: radio frequency identification (RFID) tags)
- Smart Cities: The smart city is an amalgamation of different smart domains like smart homes, smart transportation, and smart surveillance which aim is to facilitate for residents in the city, to have quality, decent life style [8,11,12,14].

- IoT in Healthcare: IoT in healthcare encompasses various devices (e.g.: heart rate monitors, ventilators, pulse oximetry monitors, etc.) that are used for patient treatment, disease diagnosis, remote health monitoring, and emergency patient care [19,20,22,23].
- Smart Grid: This is known as the next-generation power grid which emerged as a replacement for outdated power systems in the 21st century. It is integrated with advanced communication and computing capabilities that assist in controlling and managing energy resources [11,12,14,21].
- Smart vehicles: This is a novel research area that the automotive industry is currently focused on, which is developing an automobile capable of driving itself powered by electricity or other environmentally safe power sources [32].
- Smart farming: This technology can change the concept of farming as remote farming locations can be monitored and livestock can be tracked whereas this area has not been adequately researched yet [13,16].

## 3  Green Internet of Things

Although the IoT has so many problems, such as security and privacy, interoperability issues [20,23-25], energy usage will be the most critical obstacle we will face in implementing the IoT. As the number of IoT devices such as RFIDs, sensors, actuators and mobile devices connected to the Internet has risen rapidly [29], energy needs will also grow. If the billions of IoT devices are working constantly it will require massive amounts of energy on a daily basis and it will generate a large volume of data that will magnify the energy consumption. For transportation of this data and storage also increases the energy requirement. The fact that we are short of the traditional form of energy sources is also deepening the crisis. As a side effect of this massive energy consumption, it will increase the emission of carbon dioxide ($CO_2$) to the environment without control. To solve these problems GIoT is proposed [14,30,33]. It can also be described as the energy-efficient IoT procedures (hardware, software, and policy-based) that facilitate the reduction of greenhouse effect [10,14]. Fig 2 showcase the ecosystem of GIoT.

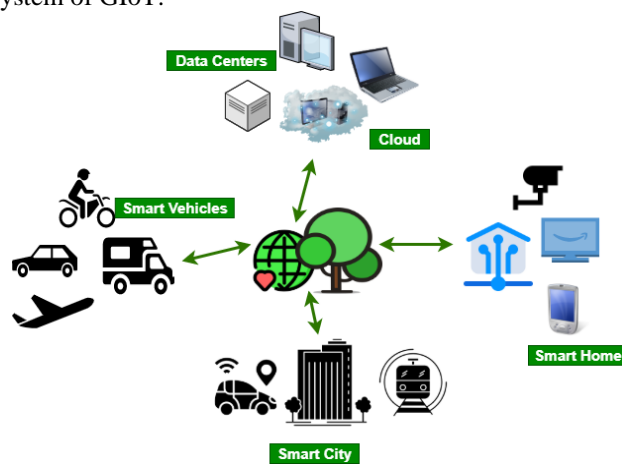

**Fig. 2.** Green Internet of Things

Before moving into the GIoT approaches, readers need to have a thorough understanding of what is meant by Green Computing (GC) as GIoT is fundamentally based on the GC techniques. GC or most popularly known as Green IT (GIT) is the study and practice of environmental sustainability computing or IT. It encompasses the study and the practice of design, manufacture, use, and the disposal of computing components efficiently and effectively with minimal effect or no effect on the environment and it consists of four phases that assist in adapting to the green practices.

- Green Use: This focuses on reducing energy consumption and promotes the sustainable use of computers and other information systems.
- Green Disposal: This focuses on the refurbishing and re-use of obsolete computers and the recycling of unwanted computer items.
- Green Design: This focuses on designing computer components that are energy efficient and environmental friendly.
- Green Manufacturing: This focuses on the development of electronic parts, digital devices with low impact or no impact on the environment.

It is noted that the eco-friendly and energy-efficiency are the two unique features of this GIoT. These characteristics are accomplished by incorporating both hardware, software and policy based energy-efficient procedures and techniques that help in minimizing energy consumption, $CO_2$ emission, and the greenhouse effect [1,3,12]. Most IoT devices are not optimized for energy efficiency, hence they waste energy when the devices are active even they are not required to be active all the time. Due to this massive energy consumption and wastage, in GIoT, it is ensured that the IoT device is ON only when it is required and idle or OFF when not required. GIoT focuses on the smart operation of devices with a decrease in energy waste. Proper energy efficient ventilation for the heat generated from servers and data centers, smart energy-conserving techniques are various strategies to conserve energy by implementing GIoT. Several key green technologies, such as green RFID, green sensing networks, and green cloud computing have been implemented to achieve GIoT. RFID is a tiny compact electronic device that contains a variety of RFID tags and small tag readers [3,4]. It basically stores data about the objects to which they are connected. The transmission range of RFID systems, in general, is a few meters. There are two types of RFID tags which known as passive and active tags. The active tags have batteries to transmit their own signal continuously, while there is no battery for the passive tags. The passive tags need to harvest energy instead of an onboard battery. Another main technology for allowing GIoT is the green wireless sensor network (WSN). A large number of sensor nodes with minimal power and storage space are used in the wireless sensor networks (WSN) [1,3,4]. Cloud computing is fundamentally based on virtualization processes, which aims to reduce energy consumption compared to having multiple servers in the data centers. Green cloud computing encompasses various policies for making the cloud more energy efficient. Following we discuss the key GIoT techniques [33-35].

- Green Internet Technology
  Green Internet Technologies require special hardware and software which are specifically designed to consume less energy without reducing performance. That includes gateways, routing devices and communication protocols etc.
- Green RFID Tags
  Active RFID tags have built-in batteries for continuously transmitting their own signal while passive RFID tags don't have an active battery source. Reducing the size of an RFID tag can help in reducing the amount of non-degradable material and there are various strategies have been proposed to reduce the energy consumption of RFID tags. Interested readers are encouraged to refer [1,34,35], for better understanding about Green RFID and associated technologies.
- Green Wireless Sensor Network
  Green WSN can be achieved by green energy conservation techniques, radio optimization techniques, and green routing techniques which leads to a reduction of mobility energy consumption in WSNs. Smart data algorithms can also be devised to reduce storage capacity as well as the size of the data content passing in the WSNs. Also to reduce the energy consumption sensor nodes in the WSN can be activated only when it is necessary.
- Green Cloud Computing
  In green cloud computing, hardware, and software are used in such a way to reduce energy consumption. Additional policies are applied to make the underlying processes more energy-efficient [1,33].
- Green Data Centers
  Data centers are responsible for storing, managing, processing, and disseminating all types of data as well as applications. Data centers should be designed in a way to use renewable energy resources. On the other hand energy-efficient ventilation techniques, energy-efficient communication protocols needed to be devised to reduce energy consumption.

In addition to the above-mentioned GIoT technologies, by modifying transmitting power (to the minimum required level) and carefully applying algorithms to design effective communication protocols, energy efficiency can be wisely improved. Also, activity scheduling, the purpose of which is to move certain nodes to low-power service (sleeping) mode, can further strengthen the energy efficiency of the networks such that only a subset of connected nodes remain active while the network is still working [3,4]. Further, there are huge concerns over the toxic pollution and E-waste that is generated as a result of this IoT ecosystem. It places new stress on achieving a sustainable eco-friendly world. There is a growing desire to shift into GIoT as it concerns and cares about the entire pervasive IoT ecosystem. In order to do this, a range of steps need to be taken to reduce $CO_2$ emission, E-waste, and should encourage device manufactures and end users for devising effective energy efficient techniques [29].

### 3.1 GIoT Approaches

The intention of this subsection is to provide readers with a brief understanding of recent models and techniques that have been developed and proposed towards achieving GIoT. We have categorized them based on the devised GIoT approach that is Hardware-Based (HB), Software-Based (SB) and Policy Based (PB).

Table 1. GIoT approaches

| Reference | Technology | Type | Description |
| --- | --- | --- | --- |
| [5] | GIoT network | SB | To extend the life expectancy of the IoT networks, energy efficient scheme is proposed in this study. |
| [6] | GIoT network | SB | An energy management scheme for IoT is introduced in this study. |
| [9] | Wireless sensor network assisted IoT network | SB | Energy efficient data routing protocol for data transferring is introduced and experimented in this study. |
| [42] | Virtualization framework for energy efficient IoT networks | HB | An energy efficient cloud computing platform for IoT is introduced in this study. |
| [43] | Controlling greenhouse effect for precision agriculture | HB/SB | IoT and cloud based system for precision agriculture is introduced in this study. |
| [44] | IoT energy management | SB | IoT energy management scheme is proposed in this study. |
| [45] | Data center | SB | A methodology for context aware sever allocation for energy efficient data center is introduced through this study. |
| [46] | Smart home automation | PB/SB | In this study authors propose various strategies to track different types of energy consumption parameters and reduce the energy wastage in smart home environment. |
| [47] | IoT sensors | SB | A method for improving energy efficiency in IoT sensors is proposed in this study. |

## 4 Challenges and Future Directions

In GIoT, there are several problems associated with transforming from IoT to GIoT. It can be based on different parameters like hardware-based, software-based, routing algorithm-based, policy-based and etc. Hardware-based can be processors, sensors, servers, ICs, RFID devices, etc. While software-based can be cloud-based, virtualization, data centers, etc. Policy-based can be smart metering systems, prediction of us-age of energy, and so on [7,12,14]. GIoT technology is currently in its infancy, but immense research activities are underway to achieve green technology and keep the environment safe. As of now, there are many difficulties and issues that have to be tackled with the urge. Following, we discuss the key challenges that are blocking the way towards achieving GIoT.

- Universal GIoT Architecture for IoT: Various vendors and standardization organizations are trying to allow links between heterogeneous networks and IoT devices with huge varieties for introducing an energy efficient architecture that can apply universally for pervasive IoT ecosystem. But due to the heterogeneity of devices and networks it has become a tedious task.
- Green Infrastructure: Offering energy-efficient infrastructure is taken into consideration as a vital issue. However, due to the complexity of deploying significantly new infrastructure, this location of research is less focused and requires more attention.
- Green Security and QoS (Quality of Service): Execution of security encryption algorithms puts the extra load on IoT devices which cause to the consumption of high energy and power. In the case of GIoT, safety and security are becoming high priority [24]. Along with achieving security for GIoT, we also need to check for solutions that gives the optimal required QoS for its users.
- GIoT Applications: This is a less focused research area whereas application layer services can be made more energy-efficient. This can be achieved through incorporating various methods when designing applications such as web applications ( e.g.: Blackle energy-saving internet search uses a black background as its search background )
- Reliability and the Content awareness: Reliability and the context-awareness should be enhanced for green IoT energy consumption models, as it leads to the making of reliable and trustworthy GIoT solutions.
- Complexity of IoT Infrastructure: It is no doubt that the IoT ecosystem comprised of a variety of complex devices. Based on the vendor and the underlying technologies devices are getting more complex and should have a proper way to reduce the complexity so more efficient energy-efficient mechanisms can be introduced.
- Green Design in Practice: As a part of the design process, when designing IoT devices, concerns like reducing energy consumption and make the devices more energy efficient with less E-waste need to be considered and incorporated. Thus it leads saves time and cost both.

### 4.1 Future Directions

It is no doubt that the quality of lives and the environment can be enhanced by GIoT, by making the related technologies and related infrastructure more environment friendly. Recent GIoT research has mainly focused on Green IoT applications and services, devising advanced energy-efficient RFIDs, energy-efficient models and planning, and localization of GIoT devices [1,4,38]. Also, it is expected that most of the IoT devices will be made in a way to recycle, again and again to reduce the toxic and hazardous materials that emit to the environment. In addition, we can expect that incorporating GIoT with enabling technologies like cloud computing [37,48], fog computing, edge computing, and blockchain will be more common with GIoT solutions as those technologies are good at providing more scalability, security, and high performance for the underlying IoT ecosystem [8,10,14,15,31].

## 5      Conclusion

Inspired by achieving a sustainable green smart world, our study provides an overview of GIoT and various integrated technologies and challenges pertaining to the GIoT. Then the future research directions and open problems regarding GIoT have been also presented. Based on our review, we noted that GIoT has the potential to offer many advantages, such as environmental sustainability and protection, end-user satisfaction in different IoT domains, minimize the harmful effects on the environment and human health. Also we noted that even though GIoT is currently in its infancy, there are lot of GIoT based research activities are being conducted towards keep the environment safe and reduce harmful effects of using IoT. As IoT constitutes the main part of digital infrastructure in the world, the benefits we can gain from adapting to green practices will be immense. We believe this study will be useful for researchers, academics, students, and other key stakeholders who are interested and working towards making a safer green world.